# The latest monthly highs suggest that the 1.5°C Paris Agreement threshold will probably be exceeded before 2028

Erhard Reschenhofer[1]

### Abstract

An attempt is made to estimate and forecast the trend of the global annual and monthly mean temperatures. The results of a conventional statistical analysis suggest that in the absence of unforeseeable events such as a sudden acceleration in the rate of warming, the 1.5°C Paris Agreement threshold could be exceeded between 2027 and 2031. However, carrying out a proper seasonal adjustment and examining the autocorrelation structure carefully, we find in a subsequent purely statistical simulation study that even the pessimistic scenario of a breach in late 2027 is inconsistent with the recent monthly highs, which means that it will probably happen much sooner.

## 1. Introduction

Bevacqua et al. (2025) observed that the first annual global mean temperature exceeding a certain global warming threshold falls typically within the first 20-year period in which the average temperature reaches the same threshold. Since in 2024 the global temperature exceeded the pre-industrial 51-year average (1850-1900) by more than 1.5 degrees C (upper limit of the Paris Agreement; see UNFCCC, 2018) for the first time, they suspected that the 20-year period that exceeds the 1.5°C limit may already have begun. Unfortunately, it will take a very long time to know for sure. Moreover, this average is not even the real quantity of interest. It is just a simple estimate of the expected global temperature in 2024. Averaging over past, present, and future values only serves to get rid of the natural variability. The inclusion of the (higher) future measurements produces a positive bias but is still necessary to compensate for the negative bias caused by the inclusion of the (lower) past values. But these two biases only cancel each other out when the trend is roughly linear. In case of an accelerating trend, the average overestimates the expected value and in the case of a slowing trend, it underestimates the expected value.

Since the upward trend of the global temperature was approximately linear and reasonably steep in the last 50 years and its variance was reasonably small, the finding of Bevacqua et al. (2025) seems plausible (see Figure 1). It would indeed be highly unlikely that the first observation exceeding a certain threshold would occur more than 10 years before or more than 10 years after the breaching of the threshold by the trend. However, the information that the Paris Agreement threshold will probably be exceeded in 10 years at the latest is still a bit vague. Of course, it would be desirable if this large time window could be narrowed down somewhat. An obvious approach is to combine past observations with climate model predictions for the future. Another approach is to use climate models to run simulations. With the help of climate model simulations, Cannon (2025) found that the Paris Agreement threshold is usually breached well before a string of unusually warm months like that from July 3023 to June 2024 occurs. In both approaches, the big question mark, however, is the reliability of the employed climate models. The situation is similar for other climatological problems. For example, you get completely different answers to the question of whether and when the Gulf Stream collapses, depending on which model is chosen, how the model's tuning parameters are specified and which statistical methods are used to estimate the model parameters (see Ditlevsen and Ditlevsen, 2023; Reschenhofer 2023, 24a,b; Van Westen et al., 2024, Reschenhofer, 2024c; Baker et al., 2025).

[1] *Retired from University of Vienna, E-mail: erhard.reschenhofer@univie.ac.at*

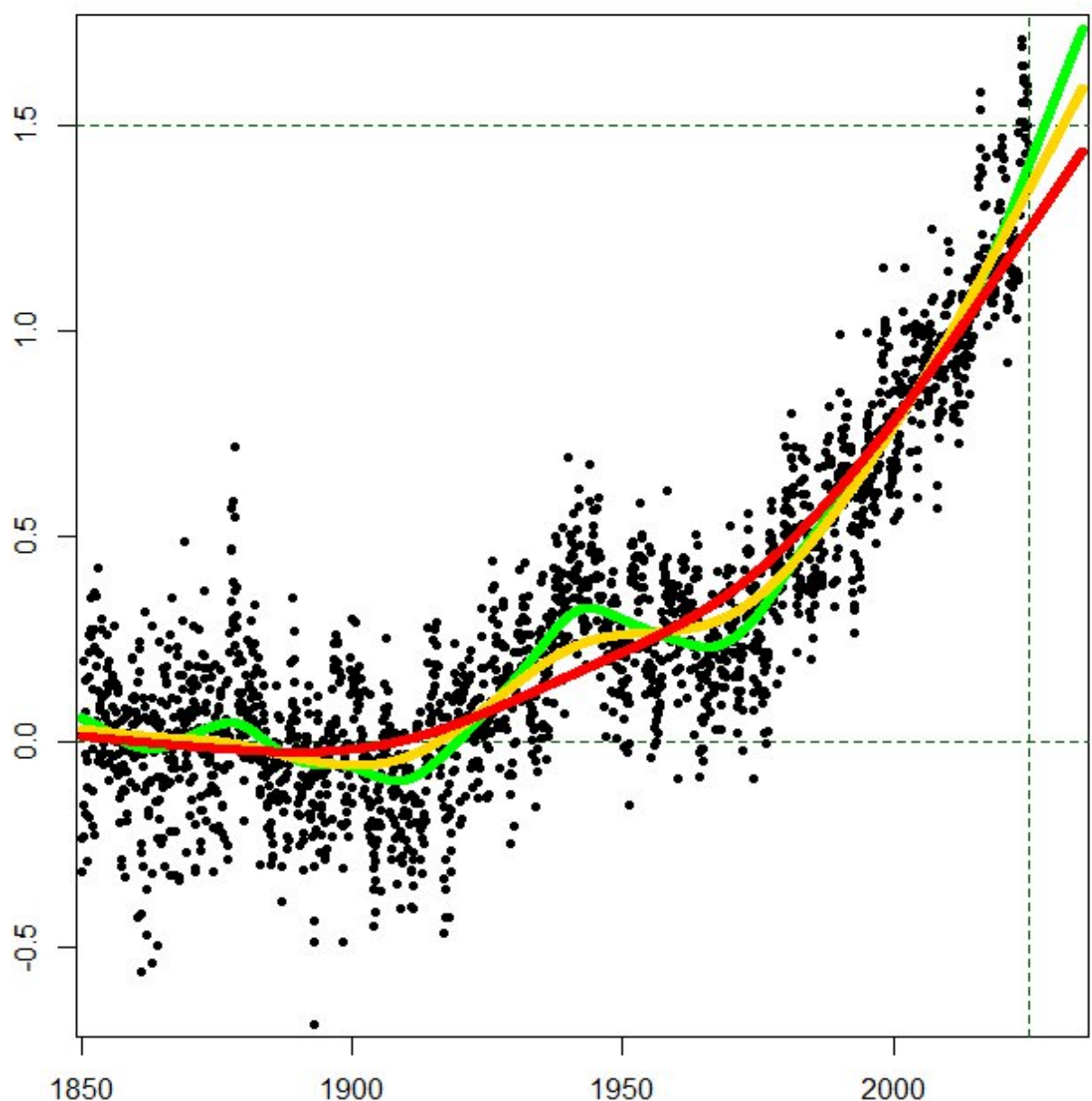

Figure 1: Extrapolation of different HP trends ($\lambda = 10{,}000{,}000$: green, $\lambda = 250{,}000{,}000$: gold, $\lambda = 2{,}500{,}000{,}000$: red) obtained from the HadCRUT5 global monthly mean temperature anomalies from January 1850 to December 2024 (relative to the average over the reference period 1850-1900)

To avoid a discussion about the quality of different climate models, a purely statistical approach is used in this paper. When it comes to forecasting, the simplest statistical models often perform better than more sophisticated models because increasing the model complexity inevitably leads to a greater variance, but not always to a significantly smaller bias, especially if the model is misspecified. The simplest models for time series focus only on trend and short-term autocorrelation. Since the latter usually fades out quickly, it can be neglected in the case of long-term forecasts. Taking a pragmatic approach to the nature of the trend, we opt for a deterministic trend simply because it is easier to forecast (for an analysis of global warming with a stochastic trend see, e.g., Mangat and Reschenhofer, 2020).

In the next section, the Hodrick-Prescott (HP) filter (Leser, 1961; Hodrick and Prescott, 1997) is used for the estimation and prediction of the deterministic trend. This estimation method has the decisive advantage over a simple average that it does not require future values that are not yet available. For the monitoring of the possibly time-varying autocorrelation, the procedure proposed by Reschenhofer (2024b) is used. All empirical results are obtained with the help of the statistical software R (R Core Team, 2024). Carrying out a proper seasonal adjustment for the monthly temperature series (as proposed by Reschenhofer, 2024b) allows a meaningful examination of the auto-correlation structure and subsequently the execution of a reasonably realistic (in the statistical sense) simulation study, the results of which are presented in Section 3. Section 4 concludes.

## 2. Estimation and prediction

Taking into account the fact that the results of the empirical analysis may depend on the choice of the temperature dataset, the reference period, and the estimation method, different choices are considered. Two datasets, two reference periods, and several values of the tuning parameter for the HP filter will be used. The first dataset is the HadCRUT5 surface temperature dataset (Morice et al., 2021), which contains global annual and monthly averages of the combined land and marine temperatures from January 1850 to December 2024. The temperatures are expressed as anomalies from the 1961-1990 period. The second dataset is the GISS Surface Temperature Analysis version 4 (GISTEMP v4; see GISTEMP Team, 2024; Lenssen et al., 2024). It contains global annual and monthly mean temperature anomalies from January 1880 to December 2024 calculated as deviations from the 1951-1980 mean. The periods 1850-1900 and 1880-1930 serve as reference periods for the two datasets. Finally, the values 10,000,000, 250,000,000, and 2,500,000,000 are used for the tuning parameter $\lambda$ of the HP filter. If instead of one long monthly time series, 12 shorter time series are considered, each with only one value per year (i.e., the annual averages or only the January averages or only the February averages or …), then the values 2,500 and 10,000 are used.

Figure 1 shows different estimates of the trend of the HadCRUT5 series obtained with the HP filter (using the R function hpfilter of the package mFilter and choosing different values for the tuning parameter $\lambda$). The HP filter estimates the trend $F_1,…,F_n$ of a time series $X_1,…,X_n$ by minimizing

$$\sum_{t=1}^{n} \left( X_t - \widehat{F}_t \right)^2 + \lambda \sum_{t=3}^{n} \left( (\widehat{F}_t - \widehat{F}_{t-1}) - (\widehat{F}_{t-1} - \widehat{F}_{t-2}) \right)^2, \tag{1}$$

where the tuning parameter $\lambda$ determines the degree of smoothing. Extrapolating the estimated trends linearly using their slopes over the last 5 years, we find that only for $\lambda = 10,000,000$ and $\lambda = 250,000,000$ the 1.5°C limit is breached in the near future but not for $\lambda = 2,500,000,000$. So the question now arises as to which value of $\lambda$ is the right one. A common practice in time series analysis is to increase the degree of smoothing until only meaningful and interpretable features remain. While there is hardly any doubt that there are still too many features in the case $\lambda = 10{,}000{,}000$, this is not so clear in the case $\lambda = 250{,}000{,}000$. Apart from the upward trend, there is still a significant feature, namely the slowdown between 1945 and 1970, which could well be a sign of an oscillation that is largely masked by the rising trend. The most likely explanation is the Atlantic Multidecadal Oscillation (AMO). However, since the AMO has large but varying amplitudes and long but varying periods, it cannot be regarded as part of the deterministic trend. A further increase in the parameter $\lambda$ therefore seems justified. Unfortunately, the complete removal of the big AMO-hump in the case $\lambda = 250{,}000{,}000$ comes at the expense of a poorer fit in other periods. A distortion at the end of the observation period would be particularly harmful, because the last few years naturally play a major role in the forecast. So despite their historical flaws, we trust the two weaker smoothings and their breaching-time forecasts December 2027 and May 2031, respectively, a little more.

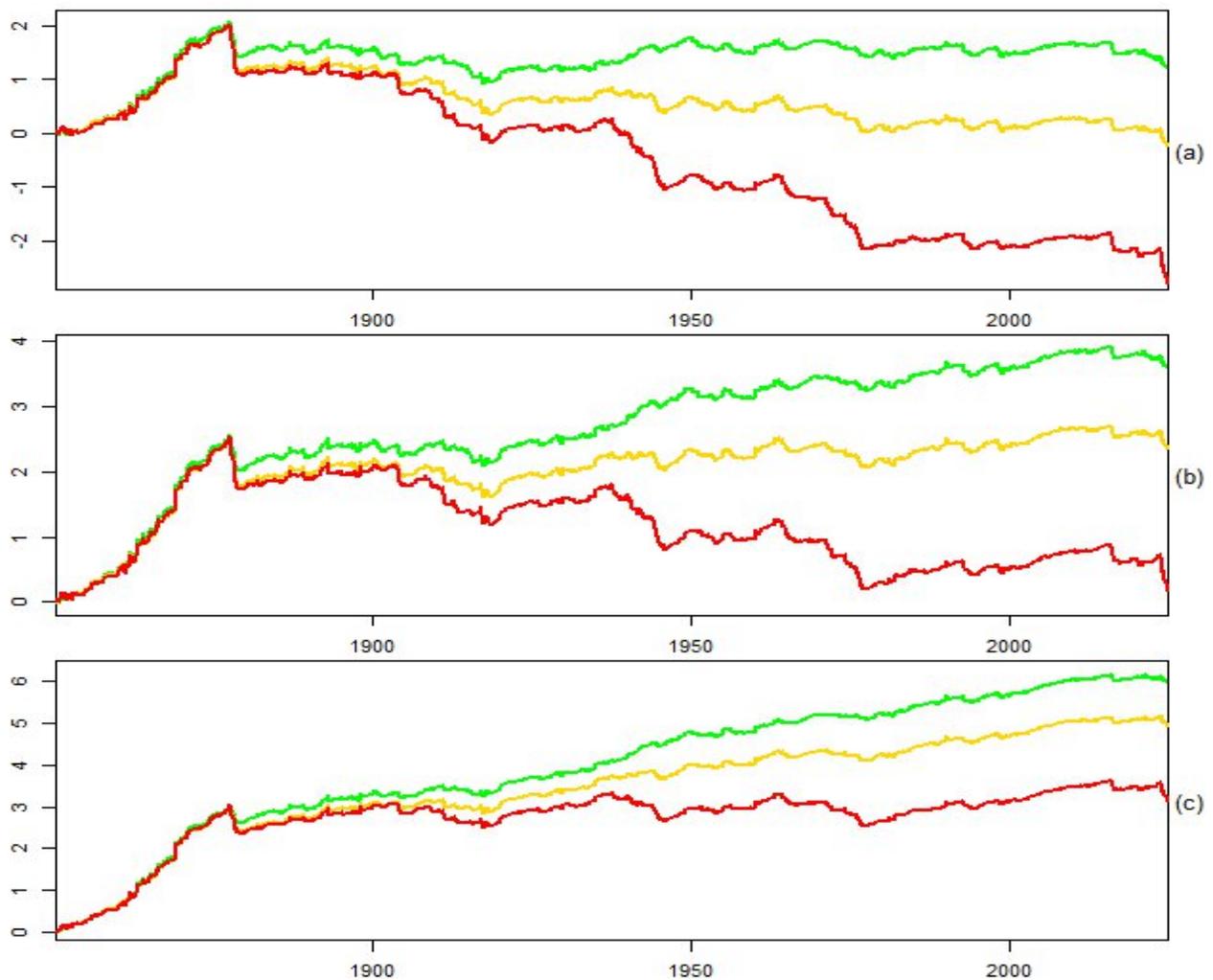

Figure 2: Monitoring the first-order autocorrelation of deviations $U_t$ from different HP trends ($\lambda = 10{,}000{,}000$: green, $\lambda = 250{,}000{,}000$: gold, $\lambda = 2{,}500{,}000{,}000$: red) obtained from the HadCRUT5 global monthly mean temperature anomalies from January 1850 to December 2024
(a) Cumulative sums $\sum \left( c\, U_t^2 - U_t U_{t-1} \right)$ with $c = 0.65$
(b) Cumulative sums $\sum \left( c\, U_t^2 - U_t U_{t-1} \right)$ with $c = 0.70$
(c) Cumulative sums $\sum \left( c\, U_t^2 - U_t U_{t-1} \right)$ with $c = 0.75$

Clearly, it may well be that the breach of the 1.5°C limit occurs earlier than December 2027. Observing that June 2024 was the twelfth month in a row with extremely high temperatures, Cannon (2025) looked for a similar pattern in climate model simulations and found that such a pattern usually occurs only after the 1.5 °C Paris Agreement threshold has already been crossed. Since the size of the autocorrelation is a decisive factor in assessing the significance of his finding, we use the procedure proposed by Reschenhofer (2024b) to monitor the autocorrelation of the trend residuals $U_1, ..., U_n$ over a long time period. Of interest to us is not only the average size but also possible changes over time, in particular whether there has been a significant increase recently. Although the procedure is very simple, it allows the assessment of the size of the autocorrelation at any point of time. The cumulative sums

$$\sum_{t=2}^{j} \left( cU_t^2 - U_t U_{t-1} \right), j = 2, ..., n \qquad (2)$$

are calculated for various values of $c$. A decrease/increase in (2) suggests that the autocorrelation is grater/less than $c$. In Figure 2, the cumulative sums are plotted for suitable values of $c$ and different HP trends. Not surprisingly, it turns out that the smoother the trend, the higher the autocorrelation. Probable values of the first-order autocorrelation of the trend residuals obtained with $\lambda = 10,000,000$, $\lambda = 250,000,000$, and $\lambda = 2,500,000,000$ are $0.65, 0.70$, and $0.75$, respectively. There are no indications of structural breaks in the slopes, hence the autocorrelation appears to be constant in each case. But this is only true if the first 30 years are excluded. In this early period, the autocorrelation was apparently much lower. However, it may well be that the data from this period are completely unreliable and should not be used at all. Perhaps there is a good reason why the time series in the second dataset only begin 30 years later.

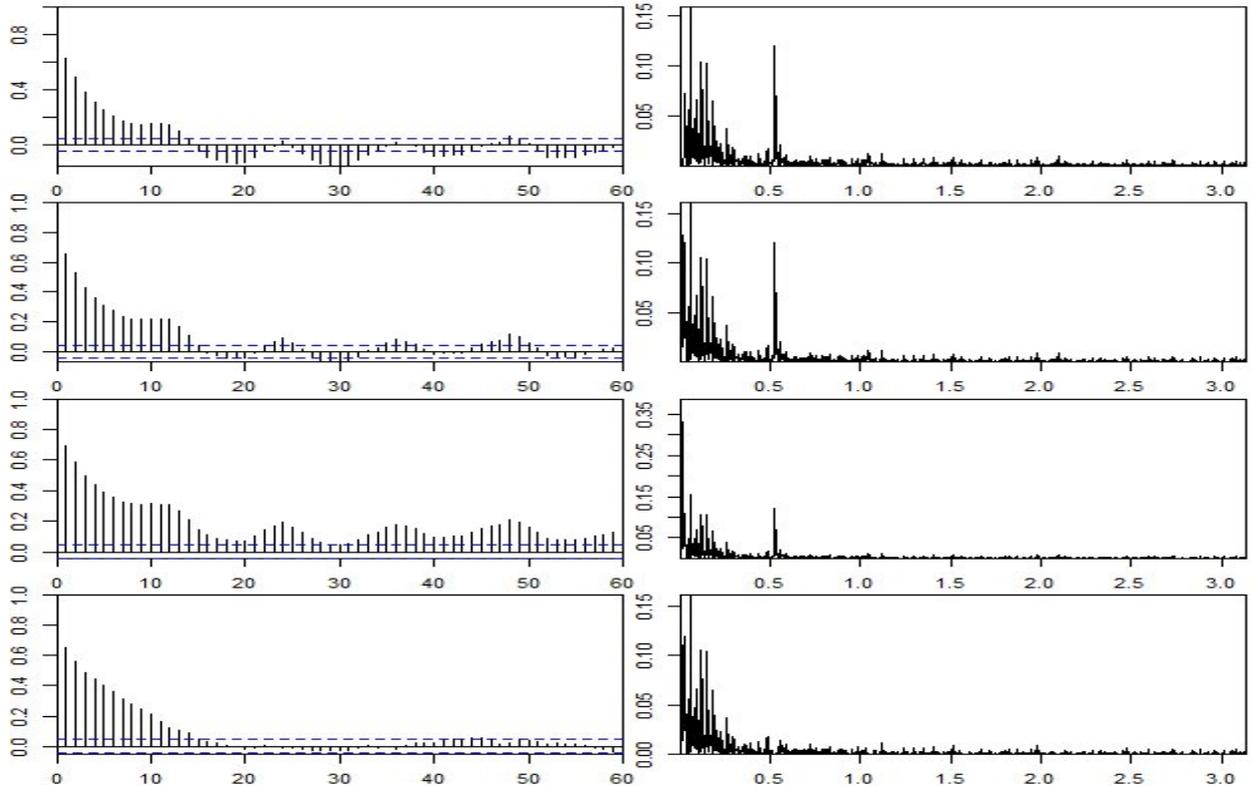

Figure 3: Sample autocorrelations and periodogram of HP trend residuals (1st row: $\lambda = 10,000,000$, 2nd row: $\lambda = 250,000,000$, 3rd row: $\lambda = 2,500,000,000$) obtained from the HadCRUT5 global monthly mean temperature anomalies from January 1880 to December 2024; 4th row: Sample autocorrelations and periodogram of properly detrended series

Despite everything that has just been said, the cluster of exceptionally high temperatures between July 1877 and September 1878 should perhaps be mentioned briefly. In each of these 15 months, the average over the previous 36 months was exceeded by 0.24 °C. In comparison, Cannon's (2025) 12 consecutive hot months (July 2023, ..., June 2024) exceeded the average of the previous 36 months by 0.26 °C. That is not a big difference. In this context, it should perhaps also not go unmentioned that the search for any conspicuous pattern is completely legitimate, but the subsequent assessment of the significance of a pattern found is somewhat more difficult. It is helpful, however, that Cannon's (2025) cluster covers exactly the last 12 months of his observation period. The potential scope for data snooping is further reduced if, in a follow-up study with more recent data, the last 12 months are no longer identical to Cannon's (2025) last 12 months.

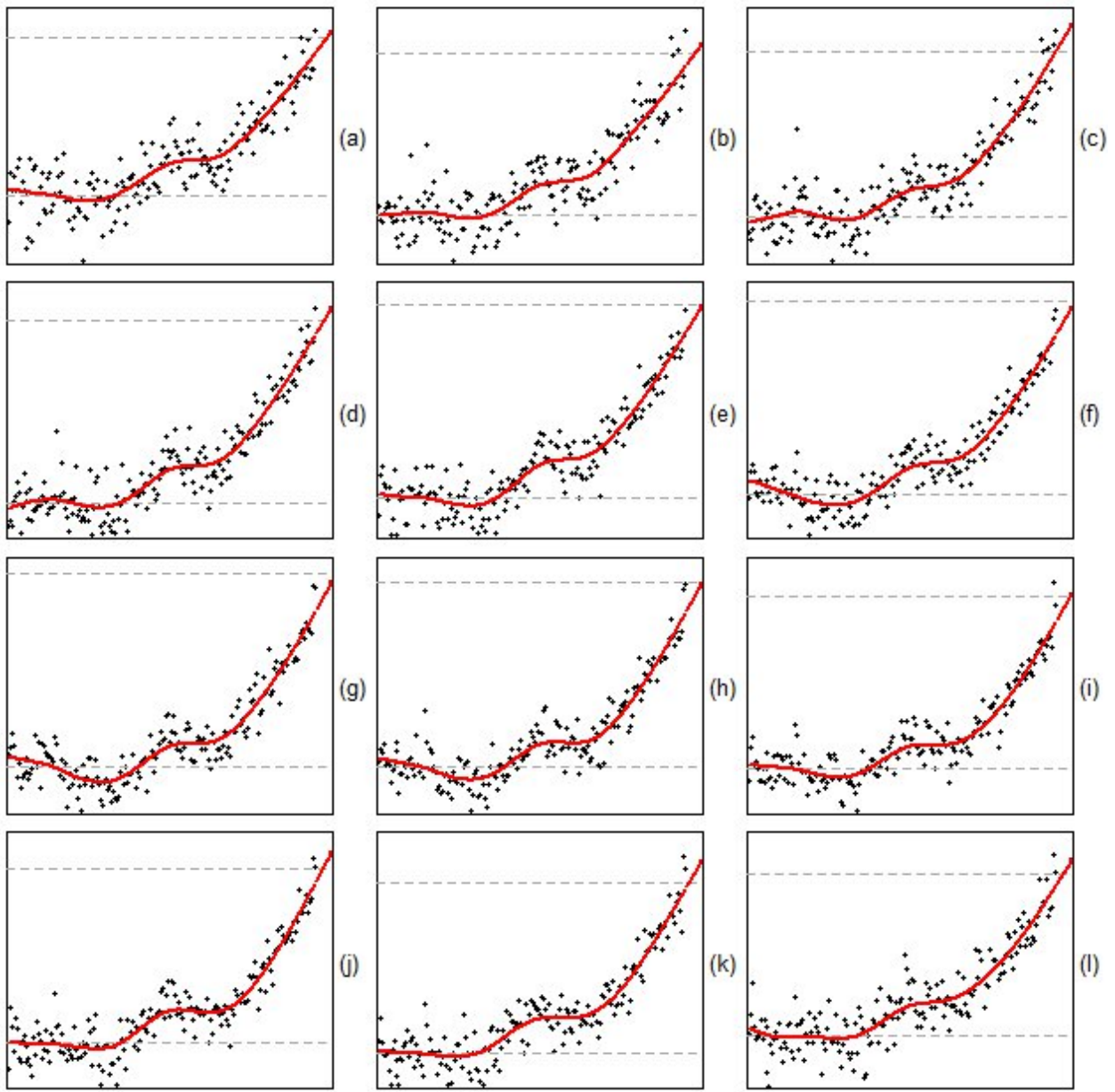

Figure 4: Estimation ($\lambda = 10,000$) and extrapolation of HP trend for HadCRUT5 dataset carried out separately for each month (a-l: January-December; dashed lines: reference levels and Paris Agreement limits = reference levels+1.5°C )

For the sake of completeness, higher-order autocorrelations and periodograms are also examined. Actually, the trend residuals obtained from series of anomalies should not exhibit any seasonality. But apparently they do (see Figure 3.a-c). Significant peaks occur in the sample autocorrelation function at lags 12, 24, 36, … and a peak in the periodogram at the seasonal frequency $\pi/6$, which can be explained by nonuniform trends. To take care of this problem, Reschenhofer (2024b) proposed to estimate the trend separately for each month. Indeed, this approach makes the persistent seasonality disappear (see Figure 3.d). The remaining short-term autocorrelation does not fall as quickly as in the case of an autoregressive process of order 1, but still fast enough to be ignored in a longer-term forecast. The 12 individual trend estimates ($\lambda = 10,000$) are shown in Figure 4. If we extrapolate the estimated trends linearly using their slopes over the last 5 years, we find that the breach of the 1.5°C Paris Agreement threshold may occur as early as 2025 (in case of the March series), but in most cases around the year 2031 (in which both the extrapolation of the annual global mean and the average of the 12 extrapolations exceed the 1.5°C limit). Using $\lambda = 2,500$ instead of $\lambda = 10,000$, we get 2030 instead of 2031.

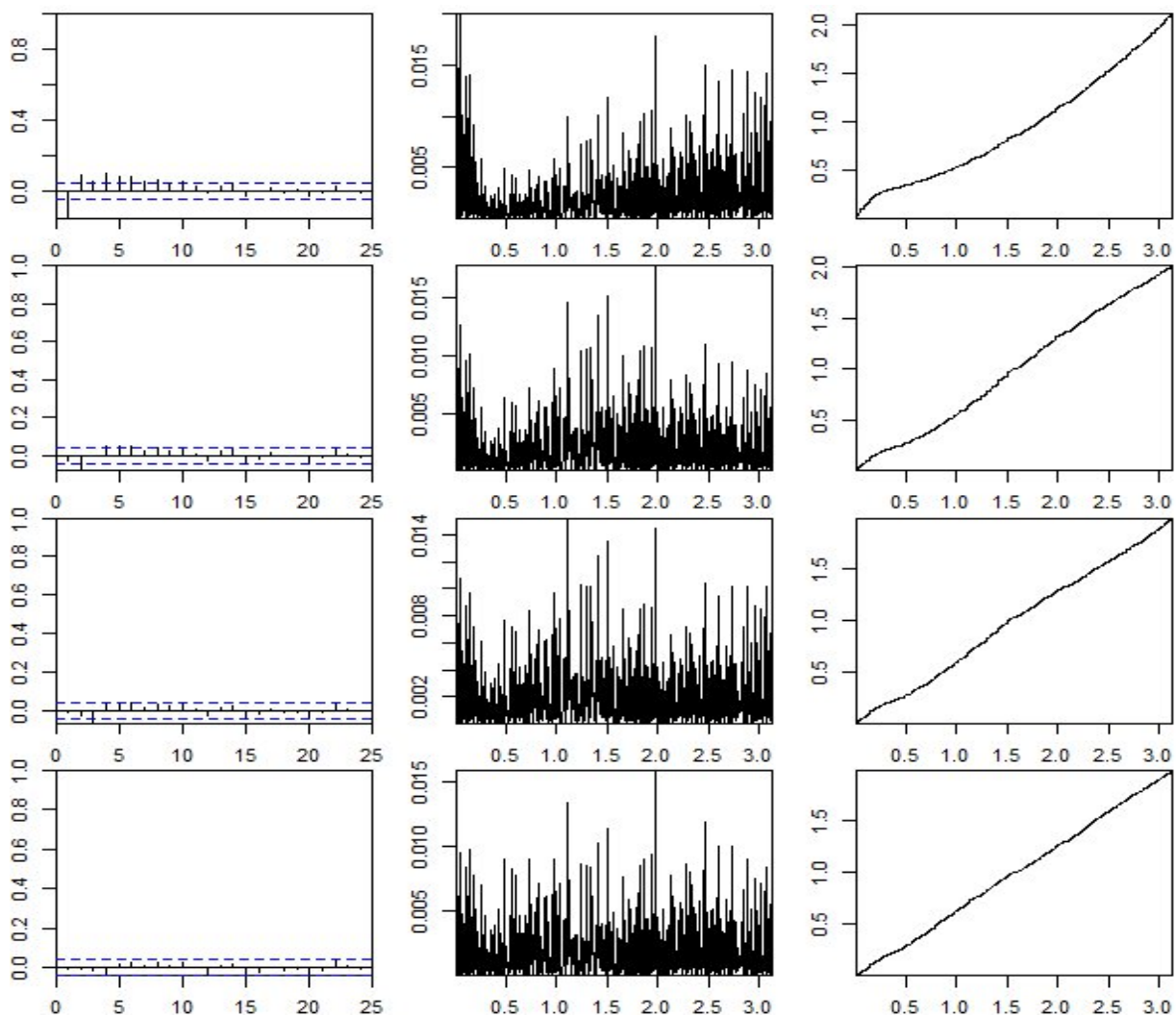

Figure 5: Sample autocorrelations, periodogram and cumulative periodogram of residuals obtained by fitting AR models of order 1 (1st row), 2 (2nd row), 3 (3rd row), and 4 (4th row) to the properly detrended monthly HadCRUT5 series

However, if we not only want to forecast but also run simulations, the trend alone is not enough. Fitting autoregressive (AR) models

$$U_t = \phi_1 U_{t-1} + \ldots + \phi_p U_{t-p} + V_t \qquad (2)$$

of increasing order $p$ to the properly detrended series, we find that the AR(4) model is the smallest model that produces residuals with a featureless periodogram

$$I(\omega_k) = \frac{1}{2\pi n}\left|\sum_{t=1}^{n} V_t e^{-i\omega_k t}\right|^2, \omega_k = 2\pi k/n, k = 1,\ldots,[n/2], \qquad (3)$$

and an almost linear cumulative periodogram (see Figure 5). For the simulations, we will therefore use the sum of the estimated trend and the AR(4) model fitted to the detrended series.

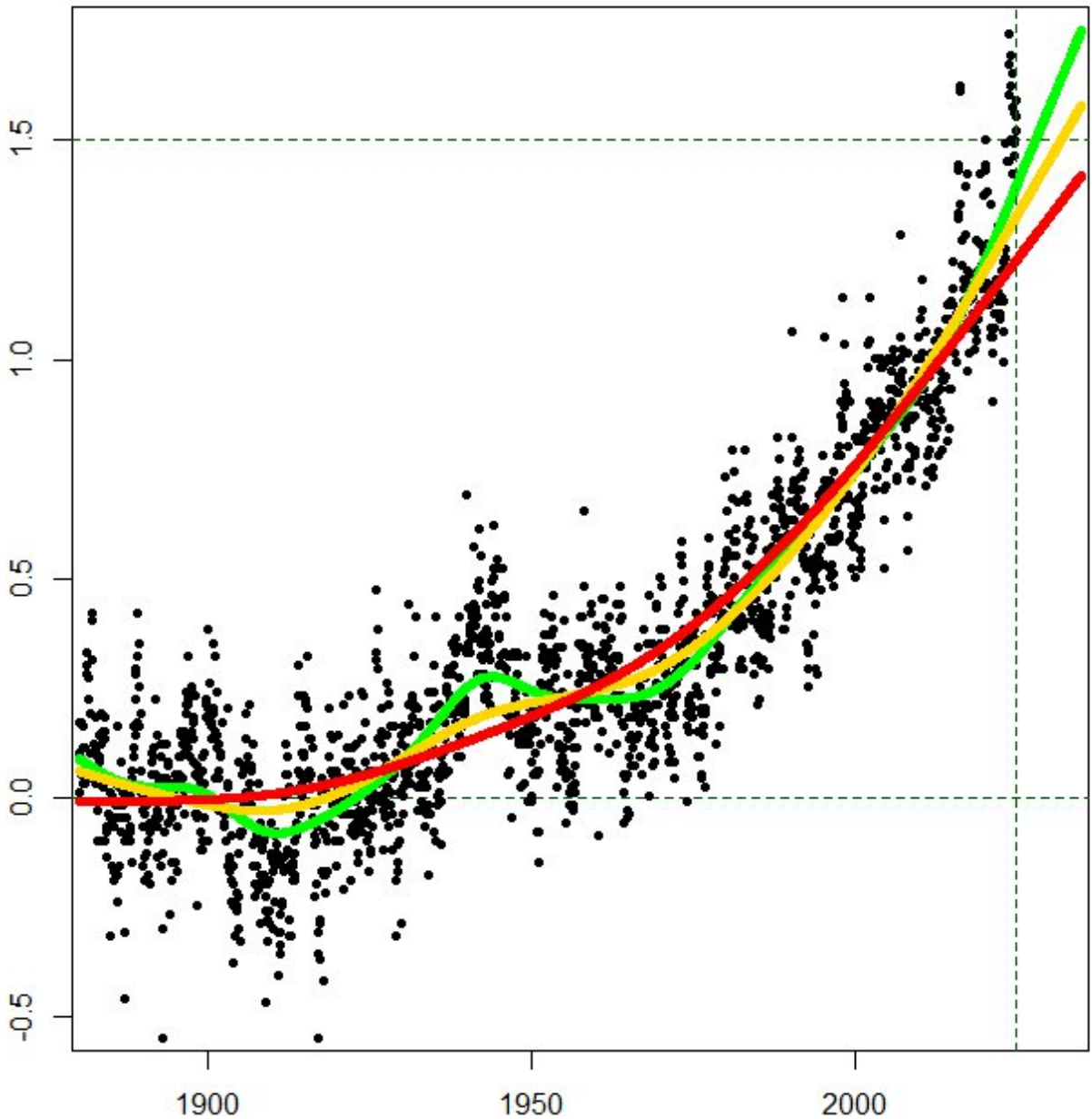

Figure 6: Extrapolation of different HP trends ($\lambda = 10{,}000{,}000$: green, $\lambda = 250{,}000{,}000$: gold, $\lambda = 2{,}500{,}000{,}000$: red) obtained from the GISTEMP v4 global monthly mean temperature anomalies from January 1880 to December 2024 (relative to the average over the reference period 1880-1930)

If we now turn to the second data set with the shorter observation period 1880-2024 and the reference period 1880-2030, we save ourselves problems with unreliable data, but by and large we get the same results as before (see Figures 6-10), which is an indication of the robustness of our results. More precisely, we find now that in the case of the 12 individual trend estimates the breach of the 1.5°C threshold may occur between 2027 (in case of the March series) and 2035 (in case of the May series), most likely around 2031 (see Figure 8). For the long monthly series, we get the forecast November 2027 if $\lambda = 10{,}000{,}000$ and October 2031 if $\lambda = 250{,}000{,}000$ (see Figure 6). Probable values of the first-order autocorrelation of the trend residuals obtained with $\lambda = 10{,}000{,}000$, $\lambda = 250{,}000{,}000$, and $\lambda = 2{,}500{,}000{,}000$ are 0.625, 0.665, and 0.71, respectively (see Figure 7).

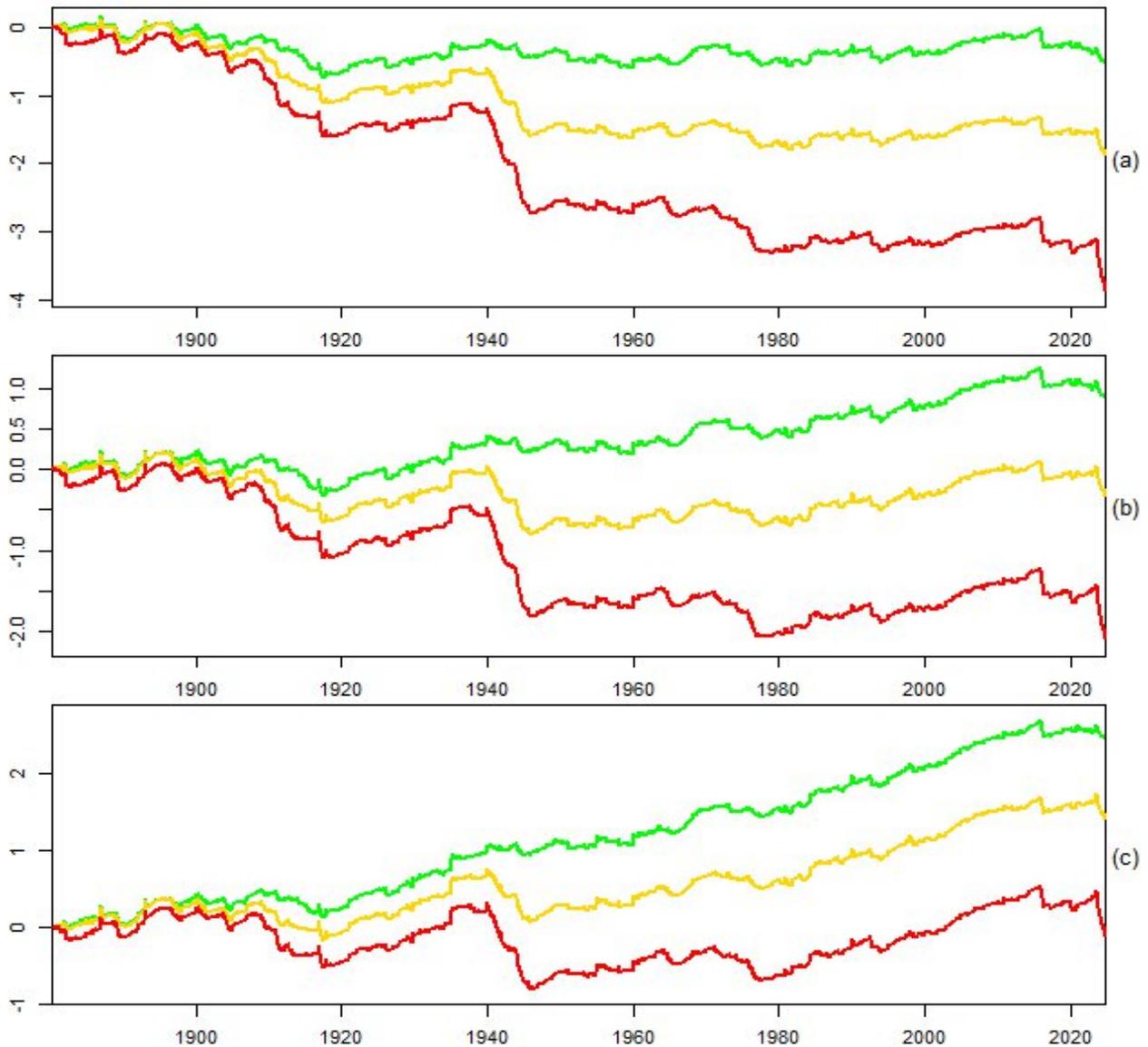

Figure 7: Monitoring the first-order autocorrelation of deviations $U_t$ from different HP trends ($\lambda = 10{,}000{,}000$: green, $\lambda = 250{,}000{,}000$: gold, $\lambda = 2{,}500{,}000{,}000$: red) obtained from the GISTEMP v4 global monthly mean temperature anomalies from January 1880 to December 2024.
(a) Cumulative sums $\sum \left( c\, U_t^2 - U_t U_{t-1} \right)$ with $c = 0.625$
(b) Cumulative sums $\sum \left( c\, U_t^2 - U_t U_{t-1} \right)$ with $c = 0.665$
(c) Cumulative sums $\sum \left( c\, U_t^2 - U_t U_{t-1} \right)$ with $c = 0.71$

### 3. Simulations

The extrapolated trends of the global annual averages exceed the 1.5°C Paris Agreement threshold in 2031 ($\lambda = 10,000$) and 2030 ($\lambda = 2,500$), respectively. We are interested in whether these possible dates for the crossing of the trend are compatible with the crossing actually observed in 2024. For this reason, we calculate in each of the two scenarios the probability that the "observed" time series will exceed the threshold for the first time in 2024 or sooner. Using the AR(2) models fitted to the respective trend residuals for the generation of 10,000 Gaussian realizations of length 185 = 175 (observation period) + 10 (extrapolation period), we get probabilities of 0.12 and 0.14, respectively, which are small but not "significant", i.e., they provide only weak evidence against a crossing as late as 2031 or 2030. In the case of the second dataset, the corresponding probabilities obtained with the help of simulations are only a little smaller (0.07 for the first scenario with the trend crossing in 2032 and $\lambda = 10,000$; 0.11 for the second scenario with the trend crossing in 2030 and $\lambda = 2,500$).

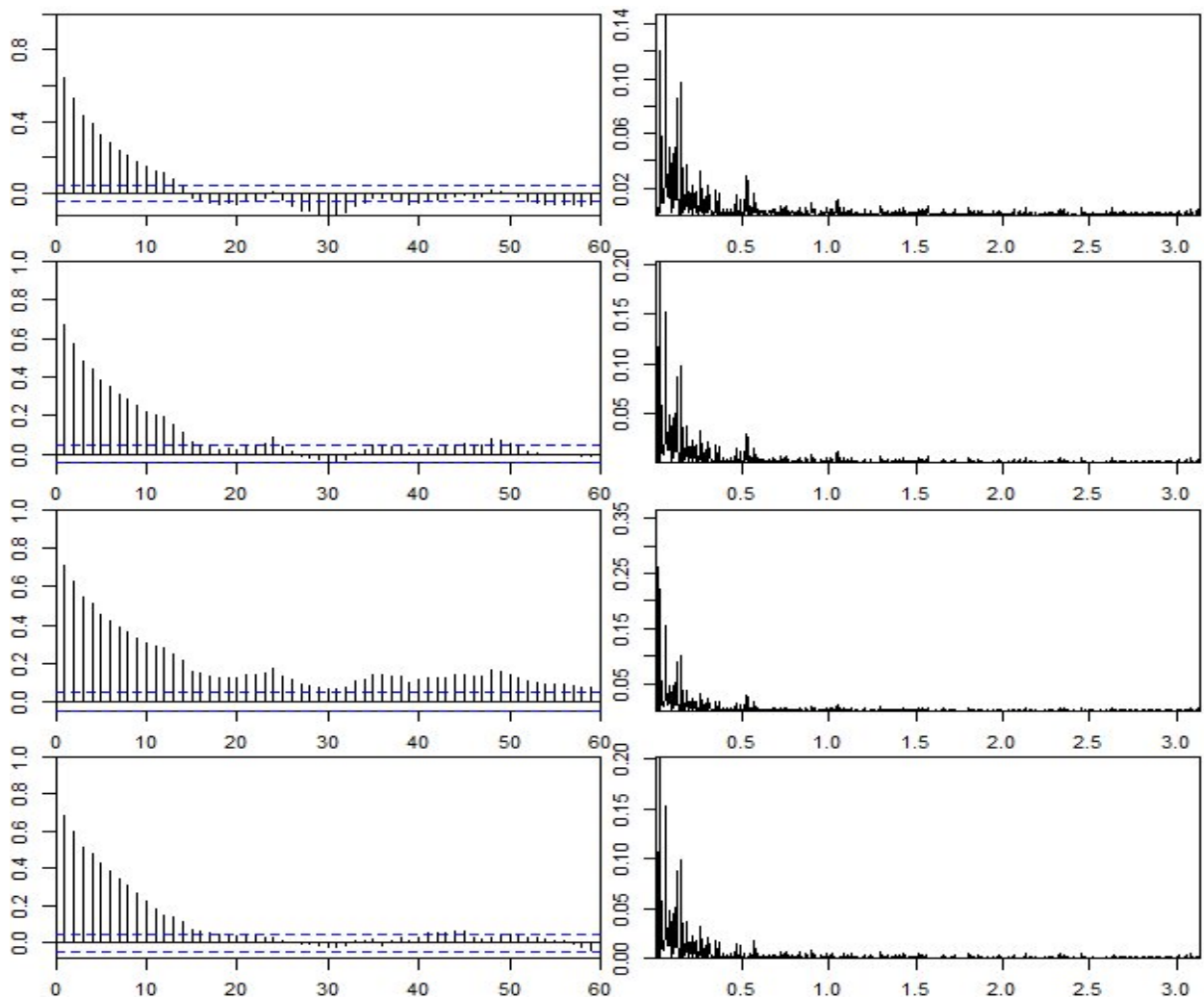

Figure 8: Sample autocorrelations and periodogram of HP trend residuals (1st row: $\lambda = 10,000,000$, 2nd row: $\lambda = 250,000,000$, 3rd row: $\lambda = 2,500,000,000$) obtained from the GISTEMP v4 global monthly mean temperature anomalies from January 1880 to December 2024; 4th row: Sample autocorrelations and periodogram of properly detrended series

Next, we look at the longer and more informative time series of global monthly averages. Of interest here is whether the number of months above the threshold observed in 2024 is consistent with one of the trends shown in Figure 1. Using the AR(4) models fitted to the respective trend residuals for the generation of 100,000 Gaussian realizations of length 2220 = 2100 (observation period) + 120 (extrapolation period), we find that even in the most pessimistic scenario with the trend crossing already in December 2027 ($\lambda = 10,000,000$), the probability of at least 8 months above the threshold is less than 5%, regardless which parameter is chosen for the seasonal adjustment ($\lambda = 2,500$ or $\lambda = 10,000$). An analogous result is obtained for the second dataset with the trends shown in Figure 6.

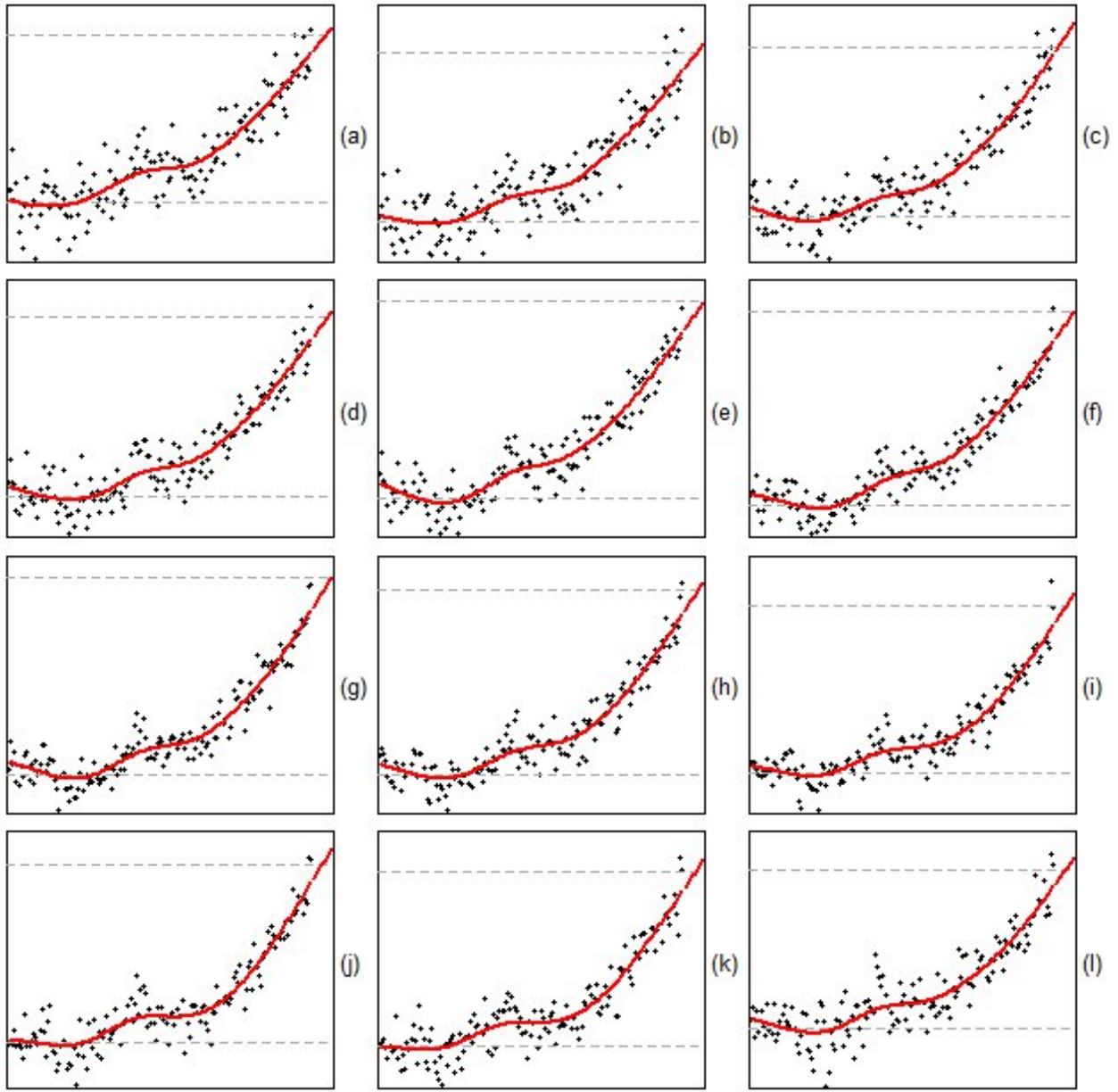

Figure 9: Estimation ($\lambda = 10,000$) and extrapolation of HP trend for GISTEMP v4 dataset carried out separately for each month (a-l: January-December; dashed lines: reference levels and Paris Agreement limits = reference levels+1.5°C )

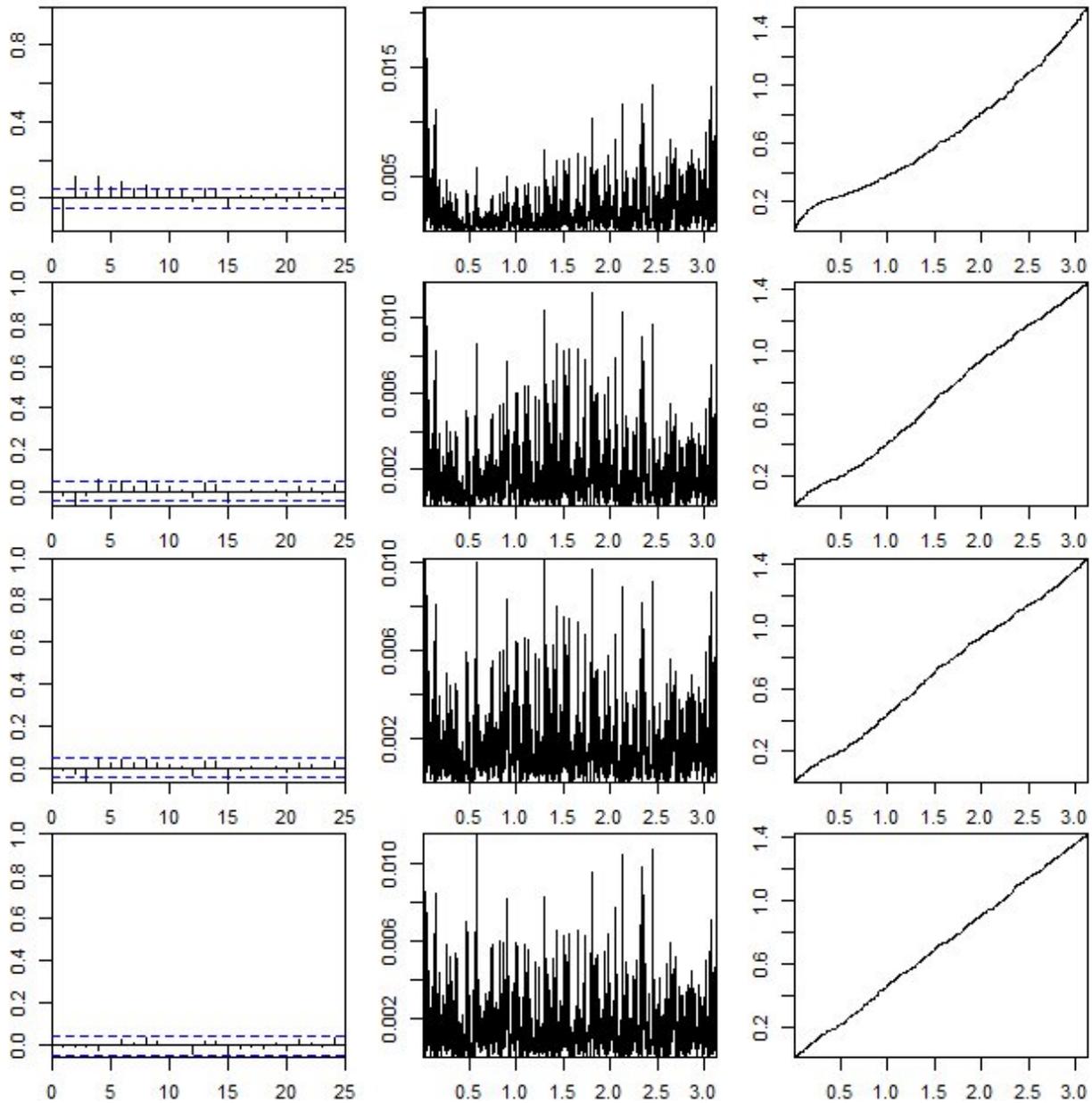

Figure 10: Sample autocorrelations, periodogram and cumulative periodogram of residuals obtained by fitting AR models of order 1 (1st row), 2 (2nd row), 3 (3rd row), and 4 (4th row) to the properly detrended monthly GISTEMP v4 series

## 4. Discussion

Inspired by patterns in historical temperature time series and encouraged by the results of climate model simulations, Bevacqua et al. (2025) suspected that the first 20-year period with an average temperature of 1.5 °C above the pre-industrial level may already have begun and Cannon (2025) even suspected that this threshold may already have been crossed. The most optimistic scenario of Bevacqua et al. (2025) is if 2024 is the first year of the crucial 20-year period and the breach happens in the middle of this period 10 years later. The best scenario of Cannon (2025) is if the breach did not occur until 2023 or 2024. In our own study, we chose a purely statistical approach and did not use any climate models. A simple estimation and extrapolation of the trend suggests that the 1.5°C Paris Agreement threshold might probably be exceeded between 2027 and 2031, earlier than in the optimistic scenario of Bevacqua et al. (2025). In contrast, a simulation study based on a more thorough examination of the statistical characteristics of the temperature time series suggests that this could well happen earlier than 2027, but has probably not already happened. However, statistical predictions are typically just smooth extrapolations and are therefore not suitable for predicting sudden and unexpected changes. So if the recent, drastic temperature rises are signs of a sudden acceleration in the rate of warming, then anything is possible. In case of such a structural break, all previous forecasts are no longer valid. Unfortunately, we can only say in retrospect whether there has been a structural break or not. The good thing, however, is that structural breaks are very rare.